# SLOTTED CSMA/CA BASED ENERGY EFFICIENT MAC PROTOCOL DESIGN IN NANONETWORKS

Suk Jin Lee[1], Hongsik Choi[2], and Sungun Kim[3]

[1]TSYS School of Computer Science, Columbus State University, Columbus, USA

[2]Information Technology, Georgia Gwinnett College, Lawrenceville, USA

[3]Department of Information and Communications Engineering, Pukyong National University, Busan, South Korea

## ABSTRACT

*Devices in a beacon-enabled network use slotted CSMA/CA to contend for channel usage. Each node in the network competes for the channels when ready to transmit data. The slotted CSMA/CA mechanism based on the super-frame structure fairly provides communication chance for each node and makes a reasonable usage of the available energy in beacon-enabled Zigbee networks. When wireless nano-sensor nodes are implanted into the target human body area for detecting disease symptoms or virus existence, each node also requires a similar characteristic in channel sharing and in the transmission of event-driven data with a short length. In this paper, we suggest a wireless network model with nano-sensor nodes for the in-body application. We propose a novel MAC protocol derived from an existing Zigbee MAC protocol scheme and analyze the performance of energy usage with variable super-frame durations and packet sizes.*

## 1. INTRODUCTION

The transition from the Internet of Things, IoT, to the Internet of Nano-Things, IoNT, stems from technological advancements of nano-metric devices based on nanoscale ElectroMagnetic, EM, communication in the THz band. As a paradigm-shifting concept in communication and network technologies, the IoNT promises to make emerging applications in Information Technology possible, applications in in-body sensors and actuator networks, environmental control of toxic agents and pollution, precise quality control of mechanical material production, and military fields [1][2][3][4].

Nanonetworks of nano-metric devices equipped with sensing and EM nano communication capabilities confine their target area within at most a few millimeters [5]. Even though graphene-based nano-antennas can be integrated into wireless nano-sensors and can support massive data exchanges in extremely high data transmission rates (i.e. Tbps) [11], there are still challenges to realize the IoNT [3]. For example, there is the possibility of the potential collisions in the shared wireless channel when the data are collected by nano-nodes which have limited power capabilities. This limitation causes nanonetworks to require a robust communication platform to guarantee an efficient TDMA energy aware MAC layer protocol [6]. Present fairly in communication chances for each nano-node, it requires a new MAC protocol which should be simple and energy-constrained.

In this paper, we propose a novel MAC protocol derived from an existing Zigbee MAC protocol for Low-Rate Personal Area Networks [7]. Actually, in a beacon-enabled Zigbee network, the MAC layer uses simple slotted CSMA/CA mechanism to contend for shared channels and to reasonably use the available energy.

In the following section, we explain related works including the application model, MAC protocol, and communication channel, which are useful for the suggested nanonetwork's applications. Second, the application model and the proposed MAC protocol are described. Third, we evaluate the proposed MAC protocol derived from the slotted CSMA/CA mechanism and compare the performance with Round Robin (RR) method. Finally, we conclude the paper.

## 2. RELATED WORKS

### 2.1. Nanonetwork Application Model

Nanonetworks are supposed to have a high number of nano-sensor nodes that are required to maintain the network connectivity for their applications, such as applications in in-body sensing, environmental control of toxic agents and pollution, or precise quality control of mechanical material production, and military fields. And nanonetworks of nano-metric devices equipped with sensing and EM nano communication capabilities limit their target area within at most a few millimeters [5].

Figure 1 depicts a simple application model of wireless nanonetworks, where our MAC protocol can be designed. The network can be composed of nano-sensor nodes, nano-micro interfaces, and a gateway regardless of any specific applications. Nano-sensor nodes are implanted into the target area for detecting specific problems (i.e., symptoms, defects, or pollutions). Each nano-sensor node will perform computations with a limited memory, and transmit small data to its master sensor node (for example, nano-coordinator) over a short range. The gateway (i.e. micro scale device) makes it possible to remotely control the entire network over the Internet.

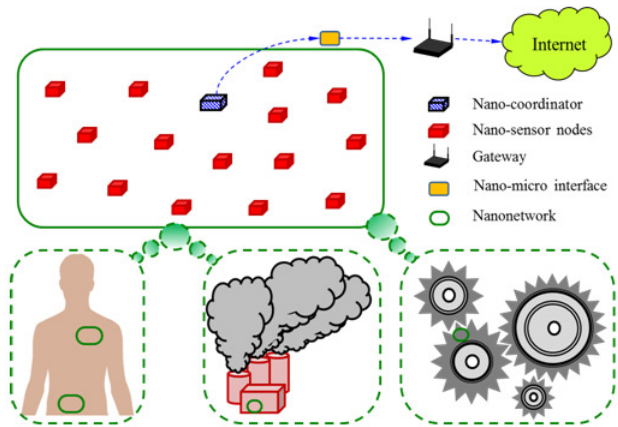

Figure 1. Application model of wireless nanonetworks

### 2.2. MAC AND ZIGBEE'S CHANNEL ACCESS MECHANISM

EM-based nano-communication network issues include the following research domains; architectures and systems for nano-communications, information theory aspects of nano-communications, communication protocols for nanonetworks or nano-sensor networks, nano-addressing and nano-coding for communications, modeling and simulation of nanonetworks, and realization of nano-applications [1][2][3][4][5][6]. In this paper, we concentrate on the domain of energy efficient MAC protocol design derived from an existing Zigbee MAC protocol for Low-Rate Personal Area Networks [7].

The nano-sensors, equipped with graphene-based nano-patch antennas, make the implementation of nano EM communications possible [8]. EM communication waves propagating in graphene-based antenna have a lower propagation speed than those propagating in metallic antenna. Nevertheless, Gbps channel capacity is available by radiating EM waves in THz frequency range. However, our suggested model for IoNT application probably does not need such a high channel capacity. With the assumption that each nano-sensor node widely dispersed in the target body area has just a role to catch specific problem only, we need to have a simple MAC protocol to communicate between nano-nodes. However, the protocol should provide a robust communication platform that guarantees the efficient TDMA energy aware MAC layer mechanism since there will be possible collisions in the shared wireless channel and the sensor nodes have a limited power (or capacity of power).

For the MAC access control among nano-nodes in a wireless electromagnetic nanonetwork, X-MAC was proposed by employing a shortened preamble approach that retains the advantages of low power listening, namely low power communication, simplicity and a decoupling of transmitter and receiver sleep schedules [9]. RIH-MAC is developed based on distributed and probabilistic schemes to create a scalable solution, which minimizes collisions and maximizes the utilization of harvested energy [10].

However, in the case of an expectedly very large number of wireless nano-sensors sharing the same channel with a very short size of data transmission (i.e., event-driven based), it is necessary to develop a new MAC protocol which has to make a reasonable usage of the available energy, and has to be simple and energy-constrained, but to fairly provide communication chance for each node. For example, nano-sensors proposed in the literature are expected to store around 800-900 of picoJoule [11] when they are fully charged. This could last for only few transmission rounds. Once the battery is drained, it might need to be recharged, for example with like piezoelectric nano-generators [12].

ZigBee is a wireless technology developed to address the needs of low-cost, low-power wireless M2M (Machine to Machine) networks and operates in unlicensed bands including 2.4 GHz, 900 MHz and 868 MHz [7]. In the star topology, the communication is established between devices and a single central controller, called the WPAN (Wireless Personal Area Network) coordinator. All devices operating on a network have unique 64 bit extended address, which can be exchanged for a short address allocated by the PAN coordinator when the device associates.

In ZigBee, the super-frame structure is an optional part of a WPAN, which has the time duration between two consecutive beacons. The super-frame duration is divided into 16 concurrent slots. The beacon is transmitted in the first slot. The remaining part of the super-frame duration can be described by the terms, CAP (Contention Access Period), CFP (Contention Free Period), and Inactive period. In a beacon-enabled network, devices use slotted CSMA/CA to contend for the channel usage. Each node in the network when it is ready to transmit data (in event-driven application) should compete for the channel. We note that a GTS (Guaranteed Time Slot) allows a device to operate on the channel within a portion of the super-frame. But we exclude this function because it is not needed for our event-driven application [7]. On the other hand, all devices sleep in the inactive part of the super-frame. Therefore, the well adjustment of the two parts (active and inactive) leads MAC channel access control to be more energy efficient for the intended application. In this paper, we adapt this slotted CSMA/CA channel access mechanism to our suggested application model.

### 2.3. TIME SPREAD ON-OFF KEYING

TS-OOK (Time Spread On-Off Keying) is a new communication technique based on the asynchronous transmission of femtosecond-long pulses among nano-sensors [13]. In TS-OOK, a single bit (logical 1 or 0) is transmitted by determining whether there exists a one-hundred-femtosecond-long pulse or not. For example, a logical 1 is transmitted by sending the femtosecond-long pulse, or a logical 0 is transmitted by remaining silent. Constant-length packets and initial preambles for the announcement of a packet transmission are used to make distinction between the silence transmission and no transmission. In TS-OOK, the pulse duration $T_p$ is much shorter than the inter-arrival time between symbols $T_s$ that is fixed for the duration of a packet, so that the receiver doesn't need to keep sensing the channel after detection of the initial preamble. In addition, the nano-sensors have the distinct symbol rate $\beta = T_s/T_p$ for different types of packets, considering the fact that the selection of optimal symbol rate $\beta_{opt} = T_s/T_p$ is still an open question [10]. Accordingly, the frequency of symbol collisions of multiple sequences in a packet is minimized [14].

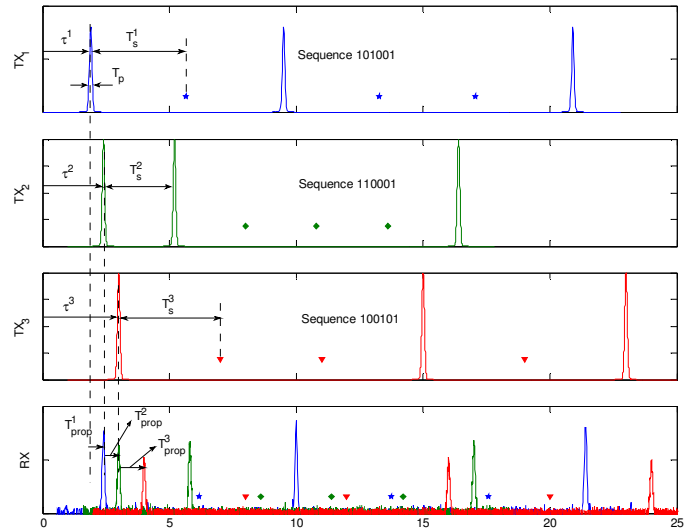

Figure 2. Exemplified TS-OOK

Figure 2 exemplifies TS-OOK with three transmitters and one receiver. Each nano-sensor intends to transmit different sequences concurrently, i.e. the sequence 101001 from $TX_1$, the sequence 110001 from $TX_2$, and the sequence 100101 from $TX_3$, to the same receiver RX. The transmitted signals undergo the propagation delay $T_{prop}$ and the molecular absorption noise. With the different transmission time $\tau$, propagation delay $T_{prop}$, and symbol rate $\beta$ for each nano-sensor, the detected signals do not overlap at the receiver side, as shown in the bottom of Figure 2. We adopt this concept for our physical layer protocol concept.

## 3. NETWORK MODEL AND PROPOSED MAC PROTOCOL DESIGN

### 3.1. NETWORK MODEL

Initially, we assume that nano-sensor nodes combine to form a nanonetwork. Nano-sensor nodes (with reduced function) are implanted into the target areas for detecting problems as intended (i.e., symptoms, defects, or pollutions), performing computation with a limited memory, and transmitting small data over a short range around 10mm, as shown in Figure 3. Whereas, the coordinator node (nano-sensor node with full function) has relatively more computational resources. Full function device acting as network coordinators will have the ability to send beacons, and to offer synchronization, communication and network join services. The coordinator node aggregates data from nano-sensor nodes (with reduced function) and sends simple data (e.g. 1 for problem detection or 0 for problem non-detection) to inform the existence of problems to the nano-micro interface. On the other hand, reduced function device can only act as end devices and are equipped with sensors and actuators, and may interact with a single coordinator only.

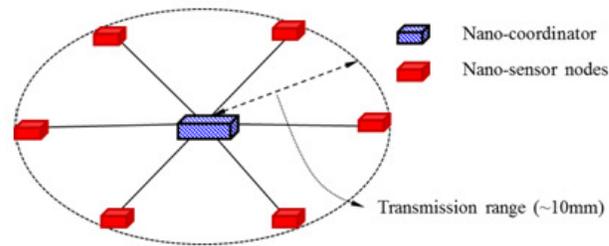

Figure 3. Network model with nano-sensor nodes and coordinator

In this paper, based on the super-frame structure given in Section 2.2, we adapt the slotted CSMA/CA channel access mechanism to our suggesting MAC model for event-driven IoNT applications. The adapted CSMA/CA channel access mechanism contributes to fairly provide communication chance for each active node and make a reasonable usage of the available energy.

### 3.2. PROPOSED MAC PROTOCOL DESIGN

This protocol is designed in three procedures; association, data transfer and disassociation. The association procedure explains that each nano-sensor becomes a member of a nanonetwork by associating with its coordinator. After channel scanning, a nano-sensor sends out the association request to the nano-coordinator using the channel available. The nano-coordinator responds to the association request by appending nano-sensors' addresses in beacon frames.

In a beacon-enabled network, there exist two types of data transfer procedure as shown in Figure 4. In the direct data transfer like Figure 4(a), a nano-sensor node finds the beacon to synchronize to the super-frame structure, and then transmit its data using slotted CSMA/CA channel access mechanism. On the other hand, the nano-coordinator indicates in the beacon that the data is pending like Figure 4(b) in the case of the indirect data transfer. For example, each nano-sensor node periodically listens to the beacon and transmits a MAC command request using slotted CSMA/CA if necessary. In the disassociation procedure, a nano-sensor node withdraws its membership by disassociating with its nano-coordinator.

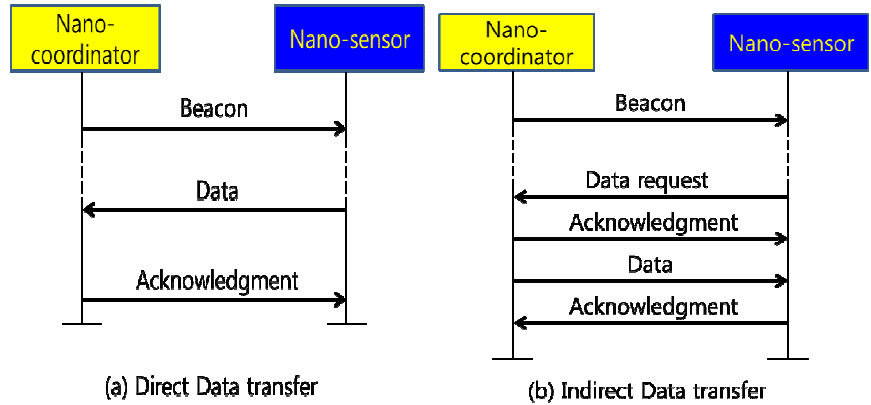

Figure 4. Data transfer procedure; (a) direct data transfer from nano-sensor to nano-coordinator, and (b) indirect data transfer from nano-coordinator to nano-sensor

Figure 5 represents the slot allocation procedure for the proposed slotted-CSMA/CA channel access mechanism. In this protocol, all concurrent nano-sensor nodes are waiting for the next time slot - once a given slot assigned. All nano-sensor nodes need to wait the beacon of the next super-frame if all the time slots of the current super-frame are already assigned.

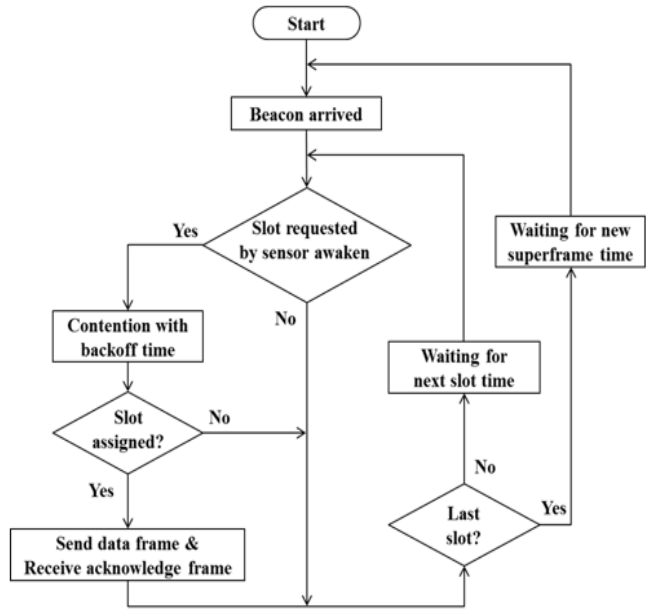

Figure 5. Slot allocation procedure for the suggested slotted-CSMA/CA channel access mechanism

The table 1 summarizes the assumed values for important terms employed for designing our MAC protocol within the framework of the super-frame in a beacon-enabled nanonetwork.

Table 1. Compare the structure of super-frame.

|  |  | 2.5 GHz | 2.5 THz |
|---|---|---|---|
| PHY frame structure |  | Sync Header (5) + PHY Header (1) + PSDU (variable) | Sync Header (5) + PHY Header (1) + PSDU (variable) |
| MAC | Data frame | MAC Header (7) + MSDU (variable) + MAC Footer (2) | MAC Header (7) + MSDU (4) + MAC Footer (2) |
|  | Beacon frame | MAC Header (7) + MSDU (3 + variable) + MAC Footer (2) | MAC Header (7) + MSDU (3 + 30) + MAC Footer (2) |
|  | Acknowledg-ement frame | MAC Header (3) + MAC Footer (2) | MAC Header (3) + MAC Footer (2) |
|  | MAC command frame | MAC Header (8) + MSDU (variable) + MAC Footer (2) | MAC Header (8) + MSDU (4) + MAC Footer (2) |

We assumed the following propositions for the structure of the proposed super-frame; the data frame is constructed by allocating a fixed-size MSDU (MAC layer Service Data Unit; 4 octets) because it is supposed for nano-sensor nodes to exchange a very short size of simple data based on event-driven. The beacon frame is constructed by a fixed-size MSDU, including super-frame specification (2 octets), pending address specification (1 octet), and address list (30 octets). The designed super-frame adopts to form a 16-bit address field, i.e. up to $2^{16}$ = 65536 nano-sensor nodes, so that it can allocate at most 15 concurrent slots during the contention access period. The MAC command frame can control association and disassociation procedure with a fixed-size MSDU (4 octets).

## 4. PERFORMANCE EVALUATION

### 4.1. ENERGY CONSUMPTIONS AND ENERGY HARVESTING MODEL

Nano-sensor nodes operating in the *THz* frequency consume low energy, whereas they maintain connectivity among nodes in the nano to millimeter communication range. Each nano-sensor node communicates with other nodes using a femtosecond-long pulse-based modulation, TS-OOK, in nanonetworks [13] [14]. Typically, the length of the pulse duration $T_p$ is much shorter than the inter-arrival time between symbols $T_s$. For considering a TS-OOK modulation scheme, we defined the length of the pulse duration $T_p$ with 100 *fs*, the inter-arrival time between symbols with 100 *ps*, and the transmission range with 10 *mm*. We assume that the energy required to transmit and receive a pulse is equal to 1 *pJ* ($E^{tx}_{pulse}$) and 0.1 *pJ* ($E^{rx}_{pulse}$), by considering ultra-low power transceivers [11][15]. The energy required handling a packet transmission and reception with *k* bits is given by [15]:

$$E^{tx}(k) = k \times w \times E^{tx}_{pulse} \qquad (1)$$

$$E^{rx}(k) = k \times E^{rx}_{pulse} = k \times E^{tx}_{pulse}/10 \qquad (2)$$

where the parameter *w* defines the probability to have a symbol 1. We assumed that the value of *w* is set to 0.5 if all symbols are generated with equal probability.

Ultra-nanocapacitor accumulates up to 800 *pJ* of energy if the piezoelectric nanogenerators realize the energy harvesting process. For common vibration sources, the time required to recharge the 95% of the ultra-nanocapacitor is 49 seconds with 50 *Hz* cycle frequency (air conditioning system) and 2419 seconds with 1 *Hz* cycle frequency (heartbeat) [11][15], as shown in Figure 6.

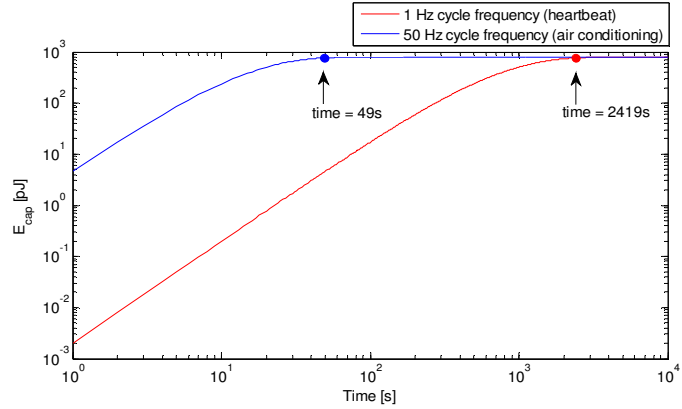

Figure 6. Accumulated energy of ultra-nanocapacitor with different cycle frequency over time

With the structure of super-frame in Table 1, and eq. (1) and (2), we can estimate energy consumption of transmission and reception for each frame. As shown in Table 2, the nano-coordinator consumes lots of energy to transmit the beacon frame rather than other frames.

Table 2. Estimated energy consumption for each assumed frame

| Property / Frame types | Packet size including PHY frame (bits) | Energy usage (*pJ*) Transmitter | Receiver |
|---|---|---|---|
| Beacon frame | 384 | 192 | 3.84 |
| Data frame | 152 | 76 | 1.52 |
| ACK Frame | 88 | 44 | 0.88 |
| MAC command frame | 160 | 80 | 1.6 |

## 4.2. PERFORMANCE ANALYSIS WITH VARIABLE SUPER-FRAME DURATIONS

We analyze the performance of the energy usage based on the framework of Zigbee standard. We change the duration of super-frames and assume that the super-frame is transmitted every 10 minutes. In the meanwhile, the nano-sensors and the nano-coordinator are able to recharge their battery up to 299.43 *pJ* with the energy harvest function, as shown in Figure 6. The performance is analyzed with the successful slot assigning rate of super-frame. If the nano-coordinator has enough energy to send the beacon frame and communicate with the active nano-sensors without draining its battery, then the super-frame is complete successfully; otherwise, the nano-coordinator passes the current super-frame and starts to send the beacon frame for the next super-frame. We assume that the nano-sensors and the nano-coordinator are fully charged at the beginning of the simulation, and the nano-coordinator has twice as its conservative energy as the regular nano-sensor.

As shown in Figure 7, we tested the performance with variable duration of super-frame. The longer the duration of super-frame, the higher the successful slot assigning rate, because the ultra-nanocapacitors are able to accumulate more energy with the extended duration. As we increased the number of concurrent slots, the successful slot assigning rate of super-frame decreased, because the nano-coordinator should consume more energy to communicate with additional nano-sensor nodes.

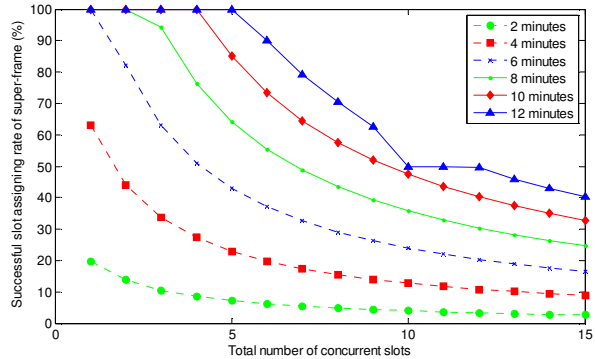

Figure 7. Successful slot assigning rate of super-frame with variable super-frame durations

### 4.3. PERFORMANCE ANALYSIS WITH VARIABLE PACKET SIZES

The performance is also analyzed with variable packet sizes. We reduced the original packet sizes gradually up to 50 percentage and checked the successful slot assigning the rate of super-frame for each duration.

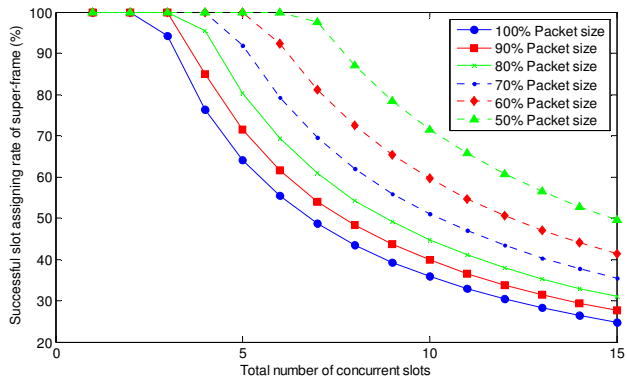

(a) Duration of super-frame: 8 minutes

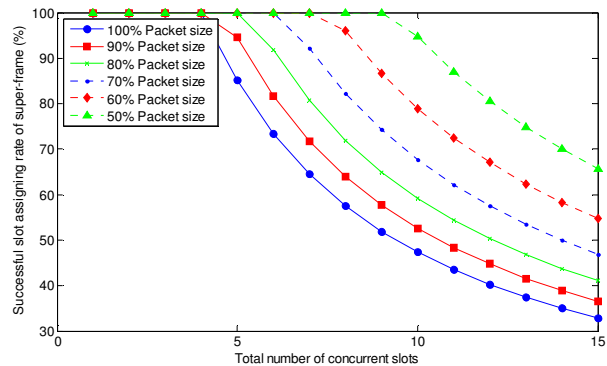

(b) Duration of super-frame: 10 minutes

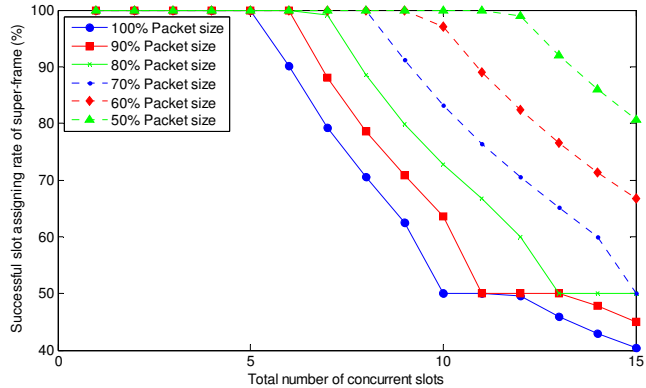

(c) Duration of super-frame: 12 minutes

Figure 8. Successful slot assigning rate of super-frame with variable packet sizes

As shown in Figure 8, the successful slot assigning the rate of super-frame is gradually increased as the packet size reduced. With the 50 percent reduced packet size and 12 minutes super-frame duration, the successful slot assigning the rate of super-frame is up to 98 percent supporting 12 concurrent slots. These results also demonstrate that the duration of super-frame and the packet size are predominant factors controlling the slotted CSMA/CA-based MAC Protocol.

### 4.4. SLOTTED CSMA/CA VS. ROUND ROBIN

We compared the possible slot usable rate for each super-frame between slotted CSMA/CA and RR (Round Robin) methods. In the given nanonetwork topology, we gradually increased the slot request rate that is the number of nano-sensor nodes to compete in a certain slot in order to send the data concurrently. For example, five nano-sensor nodes compete in the first slot to send the data in 5 percent of slot request rate. If there is a winner without collision in the slot, then the slot is usable and the corresponding nano-sensor node can send the data using this slot; otherwise, the slot is already used and then all requesting nano-sensor nodes start to compete in the next slot. The total slot number is fifteen so that they keep repeating this competition 15 times until they reach the last slot of super-frame.

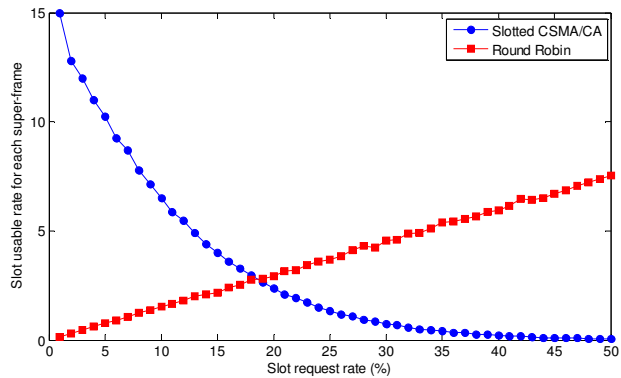

Figure 9. Comparison of possible slot usable rate between slotted CSMA/CA and RR

As the slot requesting rate increased, the possible slot usable rate of RR method increased linearly, whereas the possible slot usable rate of slotted CSMA/CA decreased. As shown in Figure 9, the slotted CSMA/CA method is superior to RR method in the event-driven application, where the slot requesting rate is less than 18 percent.

## 5. CONCLUSION AND FUTURE WORK

In this paper, we proposed a simple MAC protocol derived from the slotted CSMA/CA mechanism (based on super-frame structure) to contend for channels and to make a reasonable usage of the available energy. According to our simulation results in the aspect of energy efficiency, we experienced that the longer the duration of super-frame, the higher the successful slot assigning the rate for each super-frame. On the other hand, we noted that the successful slot assigning rate of each super-frame decreased as the number of concurrent slots requesting nodes increased. Moreover the successful slot assigning the rate of super-frame is gradually increased as the packet size reduced. These results demonstrate that the duration of super-frame and the packet size are predominant factors for controlling slotted CSMA/CA-based MAC protocol in our in-body nanonetwork application. In the future, we need to verify the proposed conceptual protocol model in the real environment to be used for the in-body application.

## Authors

**Dr. Suk Jin Lee** received the B.Eng. degree in Electronics Engineering and the M.Eng. degree in Telematics Engineering from Pukyong National University, Busan, Korea, in 2003 and 2005, respectively, and the Ph.D. degree in Electrical and Computer Engineering from Virginia Commonwealth University, Richmond, VA, in 2012. In 2007, he was a Visiting Research Scientist with the GW Center for Networks Research, George Washington University, Washington, DC. Dr. Lee worked as a faculty member in Computer Science at Texas A&M University – Texarkana for three years. He is currently an Assistant Professor at the TSYS School of Computer Science at Columbus State University, GA. His research interests include nano communication networks, network protocols, neural network, target estimate, and classification. 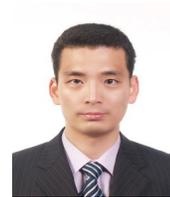

**Dr. Hongsik Choi** received the Ph.D. degree in Computer Science from George Washington University, Washington, DC, in 1996. He worked as a faculty member of several universities including Hallym University, George Washington University, Virginia Commonwealth University. He has been a member of Information Technology Discipline faculty at the Georgia Gwinnet College, Lawrenceville, GA where he is currently an Associate Professor. 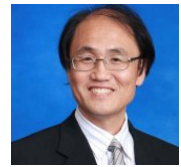

**Dr. Sungun Kim** received his BS degree from Kyungpook National University, Korea, in 1982 and his MS and PhD degrees in Computer Science from the University of Paris 7, France, in 1990 and 1993, respectively. He joined Electronics and Telecommunications Research Institute (ETRI, Korea) in 1982 and then Korea Telecom Research labs (KTRL) in 1985, where he has developed protocol testing systems for LAN, BISDN, and Intelligent Network and developed also a protocol validation tool. He has been conducting many international standardization activities in ITU-T SG7 (former) and SG15. Since 1995, he has been a professor in the Department of Information and Communications Engineering, Pukyong National University, Korea. His research interests include protocol engineering, MPLS, DWDM optical network, OTN (Optical Transport Network), wireless sensor networks and network algorithms. 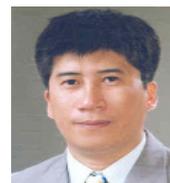